\documentclass[11pt,a4paper]{article}
\ifdefined\XeTeXversion\else\pdfoutput=1\fi
\usepackage{jcappub}
\compress
\renewcommand\afterProceedingsSpace{\vskip0pt}
\renewcommand\afterTitleSpace{\smallskip}
\renewcommand\afterRuleSpace{\medskip}
\renewcommand\afterEmailSpace{\par\vskip0pt}
\renewcommand\afterAbstractSpace{\vskip5pt}
\ifdefined\XeTeXversion\else\usepackage[T1]{fontenc}\fi
\usepackage{amsmath,amsfonts,amssymb}
\usepackage{graphicx}
\usepackage{xcolor}
\usepackage{array}
\usepackage{booktabs}
\usepackage{tabularx}
\usepackage{adjustbox}
\usepackage{url}
\usepackage{float}
\usepackage{needspace}
\usepackage{placeins}
\newcolumntype{Y}{>{\raggedright\arraybackslash}X}
\hypersetup{hypertexnames=false}
\title{Corrected detector-level reconstruction of inclined radio air showers from trigger-level ADC waveforms}

\author[a,b]{Xin Xu,}
\author[a,b,1]{Pengfei Zhang,\note{Corresponding author.}}
\author[a,b]{Xiaoyang Gong,}
\author[a,b]{Yichen Xie,}
\author[a,b]{Fufu Yang,}
\author[a,b]{Xiaosong Liu,}
\author[c,d]{and Chao Zhang}

\affiliation[a]{School of Electronic Engineering, Xidian University, No.~2 South Taibai Road, Xi'an, China}
\affiliation[b]{National Key Laboratory of Radar Detection and Sensing, Xidian University, Xi'an, China}
\affiliation[c]{School of Astronomy and Space Science, Nanjing University, 163 Xianlin Road, Nanjing 210023, China}
\affiliation[d]{Key Laboratory of Modern Astronomy and Astrophysics, Nanjing University, 163 Xianlin Road, Nanjing 210023, China}

\emailAdd{zhangpf@mail.xidian.edu.cn}

\abstract{\begingroup\fontsize{10}{12}\selectfont Autonomous radio arrays for ultra-high-energy cosmic-ray detection record air-shower radio pulses as trigger-selected analogue-to-digital converter (ADC) voltage traces. Reconstructing the primary arrival direction and energy requires accounting for noise, direction-dependent detector responses and atmospheric refraction in highly inclined showers. We develop a corrected detector-level reconstruction combining relative trigger-time correction, detector-response inversion and reference lateral-distribution function (LDF) fitting.

Waveform cross-correlation first corrects relative trigger times. Plane- and spherical-wavefront fits estimate the initial arrival direction and effective emission point, respectively. A modified voltage-domain angular-distribution fit (ADF) then refines the direction. The reconstructed source and particle axis determine the particle core and refracted radio-footprint centre. Compared with an ADF that defines viewing angles using straight source-to-station directions, the median opening-angle error is reduced by approximately 50\%. The electric field is subsequently recovered by inverting the antenna and radio-frequency-chain responses, and early--late corrections give the lateral distribution of its peak amplitudes. An empirical correction to the event-to-reference LDF peak-amplitude ratio is calibrated on 2,500 simulated electric-field showers to account for coherence loss in the 50--200 MHz band. A semi-analytic Fresnel parameterisation relates this amplitude dependence to finite-source interference. Weighted fits of the recovered lateral distribution to proton and iron ZHAireS reference LDFs are then performed, and the fitted normalisations are converted into two energy estimates using the empirical peak-amplitude correction. The final energy is their fixed equal-weight average.

Field-level and detector-level validation use 6,000 independent simulated showers. After trigger selection, angular selection and common quality cuts across three waveform treatments, 3,073 detector-level events are retained for performance evaluation. In the field-level test using simulated electric fields and simulation-input geometry, the empirical and semi-analytic corrections each reduce the root-mean-square relative energy error (RMSE) from 11.90\% without correction to 8.31\%. The detector-level reconstruction yields mean relative energy residuals of \ensuremath{-}2.30\% to \ensuremath{-}1.86\%, standard deviations of 8.46\%--8.96\%, and median opening-angle errors of $0.0315^\circ$--$0.0349^\circ$ across the different treatments. The method is intended for large-scale sparse radio arrays that detect highly inclined air showers, such as the Giant Radio Array for Neutrino Detection (GRAND).\par\endgroup}

\begin{document}
\enlargethispage{3\baselineskip}
\maketitle

\section{Introduction}
\label{sec:introduction}

Radio measurements of extensive air showers induced by ultra-high-energy cosmic rays (UHECRs) constrain the arrival direction and energy of the primary UHECRs, helping to study their origins and energy spectrum~\cite{refHuege2016,refSchroeder2017,refSchroeder2025}. Relative pulse delays and wavefront shape constrain the arrival direction. The electric-field amplitude scales approximately with electromagnetic shower energy; estimating primary energy also requires corrections for propagation geometry and emission mechanisms. Geomagnetic and charge-excess emission together produce an azimuthal asymmetry, while refractive time compression enhances the signal near the Cherenkov angle~\cite{refAugerPolarization,refLOFARPolarization,refAERAEnergy,refSchlueter2023}.

Existing methods use these relationships to reconstruct different shower parameters. Wavefront fits constrain the propagation direction and effective source distance from pulse arrival times~\cite{refLOPESWavefront,refDecoene2023}. Angular-distribution fits (ADFs) refine the direction using the amplitude distribution in viewing angle~\cite{refZhang2026E2E}. Lateral-distribution function (LDF) and radiation-energy methods estimate shower energy from the spatial signal distribution and its energy scaling~\cite{refAERAEnergy,refSchlueter2023,refGuelzow2025,refRadiationEnergy}. Two-dimensional footprint and pulse-template fits further use spatial shape and waveform information to constrain energy and shower development~\cite{refBuitink2014,refTunkaTemplate}. These methods differ in the inputs, external constraints and spatial sampling they require.

This work addresses sparse autonomous radio arrays observing highly inclined showers. The reconstruction inputs are the position, local trigger time and three-channel analogue-to-digital converter (ADC) voltage traces of each triggered station. Unlike hybrid measurements that can use particle detectors to constrain the trigger, direction or particle core, the event geometry must here be determined from the radio records themselves. The recorded traces are modified by direction-dependent antenna and radio-frequency (RF) chain responses and contain external and internal noise. Local triggering and inter-station time coincidence also determine which stations enter the event record~\cite{refNoiseRFI,refGRANDlib,refAERAL1,refTREND}. Finite bandwidth and discrete sampling further modify the measured pulse shape and peak amplitude. A small set of triggered stations must therefore constrain both local wavefront curvature and the spatial amplitude structure.

The order of reconstruction matters because an electric-field ADF requires the recovered field, but antenna-response inversion requires a direction estimate. Atmospheric refraction changes propagation delays, ray directions at the source and ground, and the radio-footprint centre relative to the particle core~\cite{refSchlueter2020CoreShift}. The local radio-arrival direction, particle-axis direction and viewing angles must therefore be distinguished. Errors in the source position and particle axis affect response inversion and shower-plane coordinates, and hence the fitted LDF amplitude.

We correct relative trigger-time offsets by waveform cross-correlation, then fit plane and spherical wavefronts to estimate the local propagation direction and effective emission point. With the source fixed, a voltage-domain ADF refines the particle axis before electric-field recovery. We replace the straight source-to-station directions used in the earlier ADF with the source-end tangents of refracted rays. The fitted source and axis determine the particle core and the central ray's ground endpoint. The latter defines the radio-footprint centre used in the LDF fit.

Field recovery builds on the three-polarisation least-squares inversion and noise weighting established in our previous collaborative work~\cite{refThreePolEField}. The previously developed detector-response and noise model extends the inversion from antenna open-circuit voltages to triggered ADC traces~\cite{refNoiseRFI}. Proton and iron reference LDFs are constructed from ZHAireS electric-field simulations using the FARSim framework~\cite{refZHAireS,refFARSim}. For energy estimation, the finite-band event-to-reference peak-amplitude ratio is calibrated as a function of source distance, atmospheric density and geomagnetic angle. A semi-analytic Fresnel parameterisation relates this dependence to coherent superposition~\cite{refChiche2024}. Both reference fits use the same reconstructed geometry, and the two energy estimates are averaged with fixed equal weights.

The reference LDFs and amplitude correction are first tested using simulated electric fields and simulation-input geometry. The full reconstruction is then evaluated using triggered ADC traces. A common event sample is used to compare noise weighting and trigger-window processing with a noise-free reference.

Section~\ref{sec:sample} describes the simulation samples, detector response and trigger selection. Section~\ref{sec:method} presents the reconstruction method, and section~\ref{sec:results} reports field-level and detector-level performance. Section~\ref{sec:discussion} compares the method with earlier work and discusses its applicability. Section~\ref{sec:conclusion} summarises the results.

\section{Simulation and detector-level event sample}
\label{sec:sample}

The validation uses simulated showers propagated through a specified detector-response model. The simulation samples, array layout and trigger settings are therefore defined before the reconstruction method, so that its inputs and evaluation conditions are explicit.

\subsection{Air-shower and site configuration}

The \emph{electric-field calibration sample} contains 2,500 ZHAireS showers evaluated before the detector response and trigger are applied. Their simulated observers form star-shaped grids: positions are sampled at different distances along radial arms at several azimuths around the shower axis, as used in the FARSim reference-LDF construction~\cite{refFARSim}. The complete calibration sample determines the final 50--200 MHz peak-amplitude correction parameters. These 2,500 calibration showers are independent of the reconstruction-validation events and do not enter any validation sample.

The \emph{fixed-array validation sample} contains 6,000 other showers. Of these, 5,235 satisfy $\theta\leq88^\circ$ and $N_{\rm fit}\geq5$, where $\theta$ is the simulation-input zenith angle and $N_{\rm fit}$ is the number of valid observer positions used in the LDF amplitude fit. Simulated electric fields and the geometry supplied to ZHAireS give the field-level results in section~\ref{sec:field_benchmark}. Propagating the same 6,000 showers through the antenna and RF-chain model described in section~\ref{sec:forward}~\cite{refNoiseRFI,refGRANDlib} gives 4,384 events passing the station-multiplicity trigger. These form the detector-level validation sample. Shower identifiers and configuration records were cross-checked to ensure that the calibration and validation samples contain no common shower.

Both samples contain proton and iron primaries simulated with QGSJET-II-04~\cite{refQGSJETII}. They span 0.001--4 EeV and predominantly inclined directions. The atmosphere follows the site-specific monthly profiles developed for Giant Radio Array for Neutrino Detection (GRAND) simulations~\cite{refGPAtmosphere}; the site coordinates, geomagnetic field and observation altitude are listed in table~\ref{tab:configuration}. 

\begin{figure}[!htbp]
  \centering
  \includegraphics[width=0.72\textwidth]{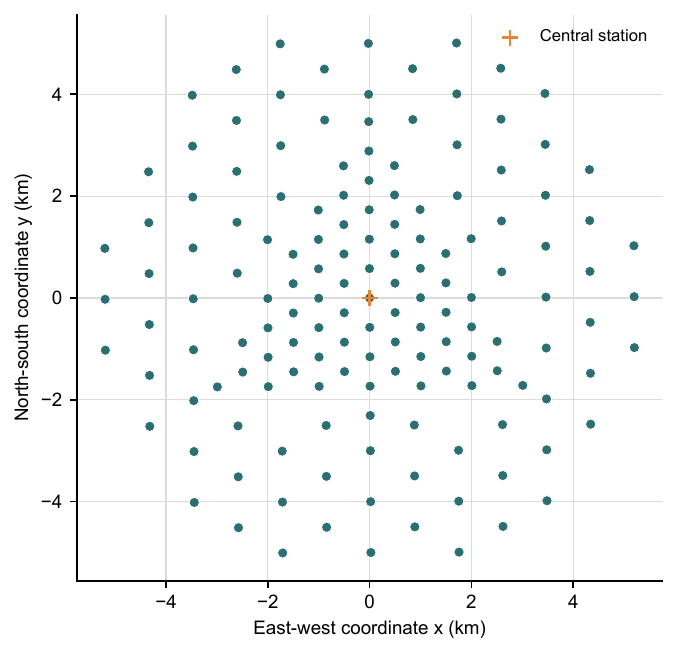}
  \caption{Fixed 148-station ground layout used for event building, timing reconstruction and lateral-distribution fitting. The orange cross marks the central station.}
  \label{fig:layout}
\end{figure}

Figure~\ref{fig:layout} shows the fixed 148-station ground layout used for detector-response propagation and event reconstruction. Unlike the star-shaped simulation grids, this layout represents the detector array on which the independent showers are observed. Its kilometre-scale spacing favours inclined showers, whose extended radio footprints can illuminate multiple stations.

\begin{table}[!htbp]
\centering
\caption{ZHAireS simulation and detector-level analysis configuration. Simulation settings are listed before detector and reconstruction settings.}
\label{tab:configuration}
\begin{adjustbox}{width=\textwidth}
\begin{tabular}{ll}
\toprule
Quantity & Setting \\
\midrule
Simulation software & ZHAireS \\
Hadronic interaction model & QGSJET-II-04 \\
Primary particles & Proton and iron \\
Observation site & $40.90^\circ$ N, $94.05^\circ$ E; 1300 m altitude \\
Generated zenith range & $45^\circ$--$89^\circ$ \\
Generated energy range & 0.001--4 EeV \\
Atmospheric model & Site-specific monthly profile \\
Geomagnetic field & $55.348~\mu$T; inclination $60.03^\circ$; declination $0.645^\circ$ \\
Other simulation settings & relative thinning $10^{-5}$; weight factor 0.06; 2 ns traces \\
Electric-field calibration & 2,500 star-shaped showers; final parameters fitted to the full sample \\
Fixed-array validation & 6,000 non-overlapping showers; 5,235 after field-level selection \\
Detector layout & 148 stations \\
ADC digitisation & signed 14 bit; $-0.9$ to $+0.9$ V input; $109.9~\mu$V per count \\
\bottomrule
\end{tabular}
\end{adjustbox}
\end{table}

\subsection{Detector response and trigger-level event selection}
\label{sec:forward}

\begin{figure}[!htbp]
  \centering
  \includegraphics[width=\textwidth]{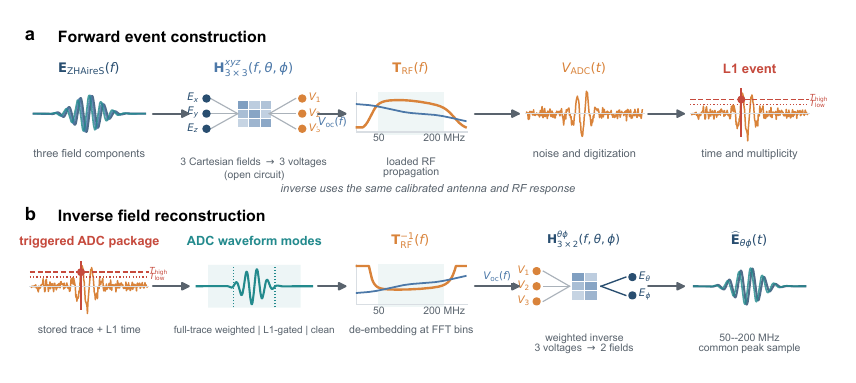}
  \caption{Forward detector response and inverse field reconstruction. Panel (a) applies the antenna effective length and circuit-modelled RF-chain response to the simulated field, followed by ADC conversion and L1 triggering. The aggregate amplitude response and noise propagation have been compared with on-site Galactic-background spectra. Panel (b) begins from the stored ADC traces: waveform processing and network de-embedding at the discrete Fourier-transform frequencies recover the open-circuit voltages, after which a three-channel complex least-squares solution reconstructs the two transverse field components.}
  \label{fig:rf_forward_inverse}
\end{figure}

Figure~\ref{fig:rf_forward_inverse}(a) shows how simulated electric fields become triggered ADC records; panel (b) shows the inverse field-recovery sequence described in section~\ref{sec:rf_inverse}. The forward calculation defines the detector-level input before reconstruction. The channel index $p\in\{x,y,z\}$ denotes the east--west, north--south and vertical antenna channels, respectively. For each channel, the direction-dependent complex vector effective length $\vec h_p$ converts the incident field to an antenna open-circuit voltage. The loaded electronics network then propagates this voltage to the ADC~\cite{refNoiseRFI,refGRANDlib}:
\begin{equation}
V_{{\rm ADC},p}(f)=G_p(f)\,[\vec h_p(f;\theta,\phi)\cdot\vec E(f)] .
\label{eq:forward_chain}
\end{equation}
Here $f$ is frequency, $(\theta,\phi)$ specifies the incident direction by its zenith and azimuth angles, and $\vec E(f)$ is the Fourier-domain incident electric field. The function $G_p(f)$ is the end-to-end complex voltage transfer function from the antenna open-circuit voltage to the ADC input. It accounts for the voltage division between the antenna impedance and receiver input impedance, together with the frequency-dependent transmission of the cascaded analogue network. The same tabulated $\vec h_p$, impedances and two-port response are used in the inverse calculation. A circuit-level model supplies the complex response. Its system-level amplitude and noise propagation have been checked against on-site Galactic-background observations: the predicted and measured ADC spectra agree to about 1 dB, and their 24-h power variations in 60--90 MHz differ by about 0.7 dB~\cite{refNoiseRFI}. The comparison tests the aggregate amplitude response and noise propagation of the circuit model.

The inverse Fourier transform is sampled at 2 ns and digitised by the modelled signed 14-bit ADC over an input range of $-0.9$ to $+0.9$ V, corresponding to $109.9~\mu$V per count. The noisy realisation additionally contains simulated Galactic and receiver noise~\cite{refNoiseRFI,refGRANDlib}. The numerical trigger values quoted below are therefore ADC-input voltages after the simulated channel gain.

The first-level (L1) station trigger uses the two-threshold pulse-shape approach developed for the Auger Engineering Radio Array (AERA)~\cite{refAERAL1}; autonomous station triggering and event selection are also demonstrated by the Tianshan Radio Experiment for Neutrino Detection (TREND)~\cite{refTREND}. For the present simulations, the condition is applied to the filtered, median-subtracted horizontal y-channel voltage. A candidate must cross the upper threshold of $6100~\mu$V and satisfy the associated lower-threshold condition at $5500~\mu$V within a 100 ns search interval following a 100 ns quiet interval; paired lower-threshold crossings may be separated by at most 25 ns. These threshold and time-window values specify the trigger settings used here. The quiet interval rejects nearby pulse trains, while the lower-threshold timing condition limits extended ringing. An event is retained when at least five stations satisfy L1. For each triggered station, the event record contains the station identifier and position, three ADC traces and the absolute L1 trigger timestamp. Waveform processing is described in section~\ref{sec:waveform_modes}, and electric-field reconstruction in section~\ref{sec:rf_inverse}.

The retained positions, trigger timestamps and ADC traces are the inputs to the reconstruction sequence in section~\ref{sec:method}.

\section{Reconstruction method}
\label{sec:method}

\subsection{Reconstruction geometry and overview}

Figure~\ref{fig:workflow} shows the reconstruction sequence. Cross-correlation-corrected times are used in the plane-wavefront (PWF) and spherical-wavefront (SWF) fits; voltage amplitudes then determine the particle axis. The fitted source and axis are used for response inversion and footprint mapping. The proton and iron reference fits use the same geometry and station coordinates, with a separate peak-amplitude correction for each reference.

\begin{figure}[!htbp]\centering
\includegraphics[width=\textwidth]{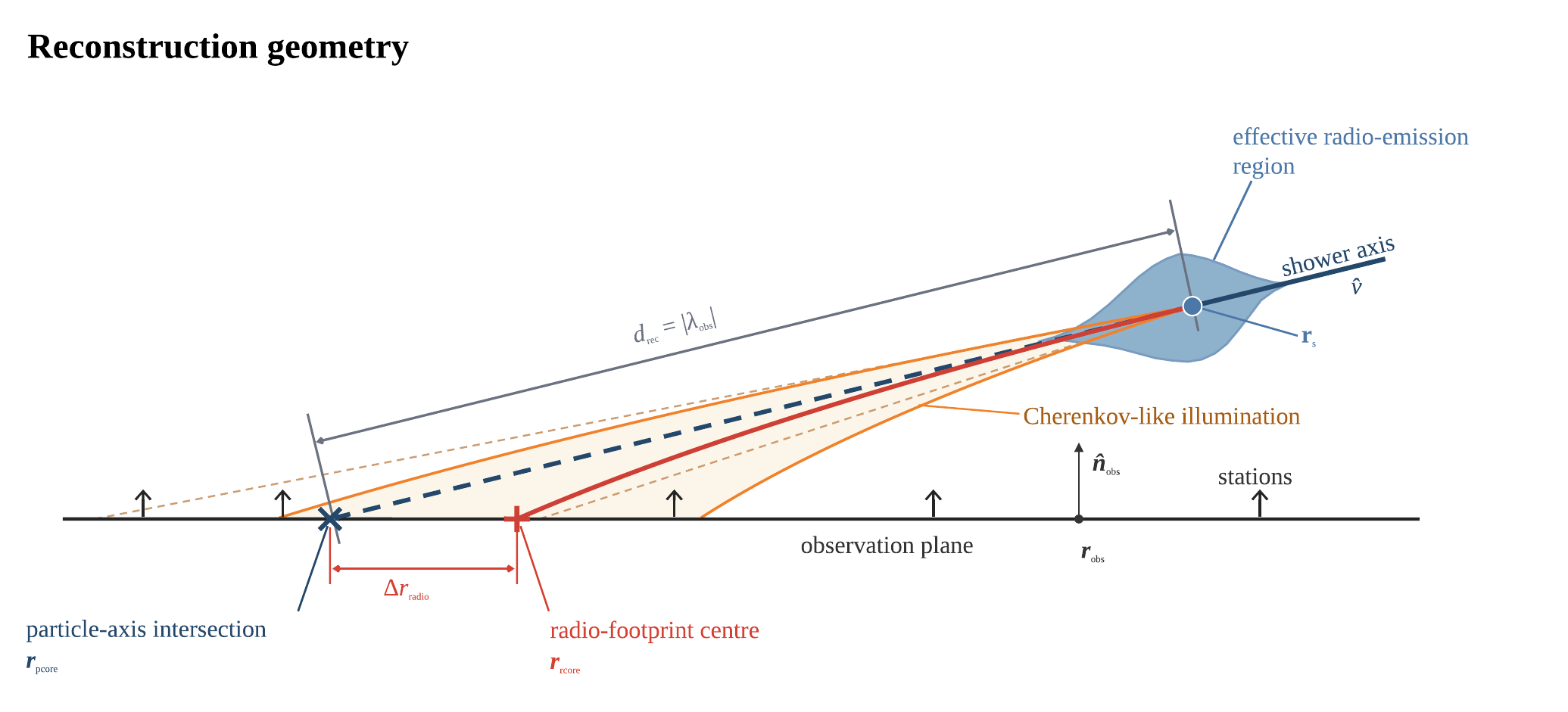}
\caption{Reconstruction geometry, adapted from Ref.~\cite{refSchlueter2020CoreShift}. The blue envelope represents the extended particle distribution and $\vec r_s$ the effective emission point. The navy dashed line is the particle axis. Orange dashed lines indicate straight-line boundaries without refraction; solid orange curves enclose the schematic refracted illumination region around the red central ray. Its ground intersection defines the radio-footprint centre, displaced from the particle core. Perpendicular end marks delimit the along-axis distance $d_{\rm rec}=|\lambda_{\rm obs}|$. The labels $\vec r_{\rm pcore}$, $\vec r_{\rm rcore}$ and $\vec r_{\rm obs}$ denote the particle core, radio-footprint centre and a reference point on the observation plane; $\hat n_{\rm obs}$ is the plane normal. The particle-axis direction $\hat v$ is $\hat k_{\rm ADF}$ in equation~\eqref{eq:final_core_distance}.}
\label{fig:pipeline}\end{figure}

Figure~\ref{fig:pipeline} defines the reconstruction geometry. Radio emission is produced over an extended region around shower maximum; its curved wavefront is approximated by a spherical wavefront centred at an effective emission point $\vec r_s$~\cite{refLOPESWavefront,refDecoene2023,refZhang2026E2E}. The particle-shower axis is the direction of the particle cascade. A refracted ray has different local directions at emission and reception, so the source-end tangent used for the ADF must be distinguished from the radio-arrival direction near the array. The particle axis intersects the observation plane at the particle core. A central ray launched from $\vec r_s$ along this axis bends through the stratified atmosphere; its ground endpoint defines the radio-footprint centre~\cite{refSchlueter2020CoreShift}.

\begin{figure}[!htbp]\centering
\begingroup\small\setlength{\fboxsep}{5pt}
\fbox{\parbox{0.91\textwidth}{\centering\textbf{Triggered event records}\\Station positions, relative L1 trigger times and three-channel ADC traces}}
\par\vspace{2pt}$\Downarrow$\par\vspace{2pt}
\begin{tabular}{@{}c@{\hspace{0.025\textwidth}}c@{}}
\fbox{\parbox[t]{0.425\textwidth}{\centering\textbf{Timing reconstruction}\\Waveform cross-correlation\\PWF direction seed; SWF source position}} &
\fbox{\parbox[t]{0.425\textwidth}{\centering\textbf{Waveform processing}\\50--200 MHz; three waveform modes\\Station voltage-envelope amplitudes}}
\end{tabular}
\par\vspace{2pt}$\Downarrow$\par\vspace{2pt}
\fbox{\parbox{0.91\textwidth}{\centering\textbf{Voltage-domain ADF with source-end ray tangents}\\Fit the particle axis and ADF shape with the SWF source fixed}}
\par\vspace{2pt}$\Downarrow$\par\vspace{2pt}
\fbox{\parbox{0.91\textwidth}{\centering\textbf{Shared event geometry}\\Particle-axis intersection; central refracted-ray endpoint}}
\par\vspace{2pt}$\Downarrow$\par\vspace{2pt}
\fbox{\parbox{0.91\textwidth}{\centering\textbf{Direction-dependent response inversion}\\$V_{\rm ADC}(f_k)\rightarrow V_{\rm oc}(f_k)\rightarrow(E_\theta,E_\phi)\rightarrow\vec E_i(t)$}}
\par\vspace{2pt}$\Downarrow$\par\vspace{2pt}
\fbox{\parbox{0.91\textwidth}{\centering\textbf{Shower-plane mapping and peak-amplitude correction}\\Early--late amplitude factor; radial distance from the radio centre\\Event-to-reference correction from reconstructed geometry}}
\par\vspace{2pt}$\Downarrow$\par\vspace{2pt}
\fbox{\parbox{0.91\textwidth}{\centering\textbf{Energy estimation and common-event evaluation}\\Weighted proton and iron LDF fits; fixed equal-weight energy average\\Quality cuts; direction and energy residuals against simulation inputs}}
\endgroup
\caption{Reconstruction from triggered ADC records to arrival direction and energy. The timing and waveform-processing branches provide the inputs to the voltage ADF. Both reference fits use the same reconstructed event geometry.}
\label{fig:workflow}\end{figure}

In a spherically stratified atmosphere, ray tracing uses the Bouguer invariant
\begin{equation}
b_{\rm ray}=n(r)r\sin\zeta,
\label{eq:ray_invariant}
\end{equation}
where $r$ is geocentric radius, $n(r)$ is the atmospheric refractive index at that radius, and $\zeta$ is the angle of the ray tangent to the local radial direction. The quantity $b_{\rm ray}$ is constant along a ray in this atmosphere. This invariant determines both the station-ray source tangents used by the ADF and the endpoint of the central ray. If $\vec r_{\rm pcore}$ is the straight-axis intersection and $\vec r_{\rm rcore}$ the central-ray intersection with the observation surface, then
\begin{equation}
\begin{aligned}
\Delta\vec r_{\rm radio}&=\vec r_{\rm rcore}-\vec r_{\rm pcore},\\
\vec r_{\rm rcore}&=\vec r_{\rm pcore}+\Delta\vec r_{\rm radio}.
\end{aligned}
\label{eq:core_shift}
\end{equation}
The refractive displacement follows the mechanism described in Ref.~\cite{refSchlueter2020CoreShift}. In this reconstruction, the same fitted source and axis determine both intersections, and the station rays and central ray use the same refractivity profile. The central-ray endpoint replaces the particle core as the LDF radial origin, accounting for the offset between the particle axis and the radio footprint.

\subsection{Waveform processing and pulse-time alignment}
\label{sec:waveform_modes}

\begin{figure}[!htbp]
  \centering
  \includegraphics[width=\textwidth]{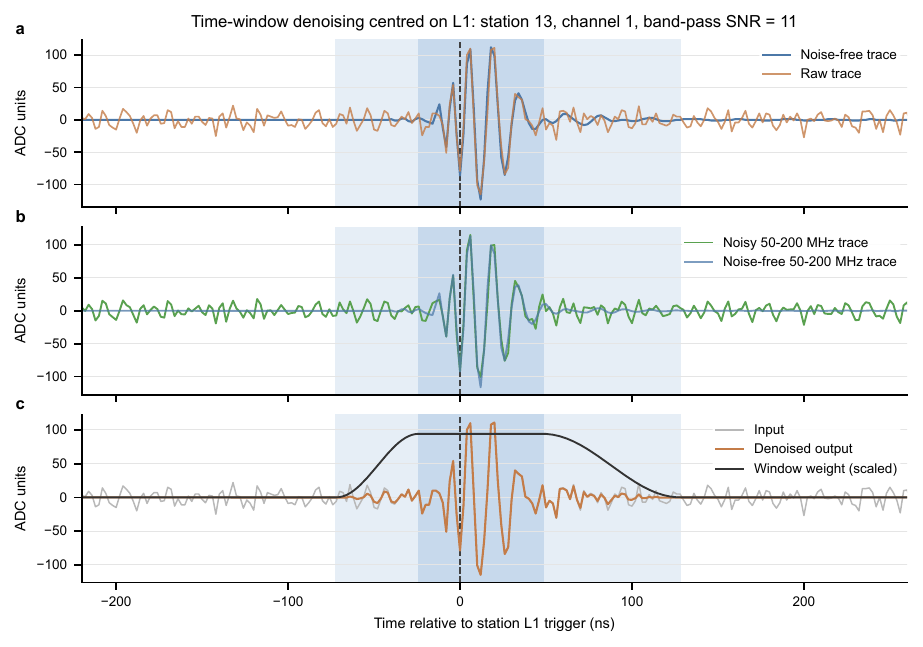}
  \caption{L1-anchored waveform processing applied before the voltage ADF and RF inversion. (a) Noisy and noise-free ADC traces. (b) The same traces after 50--200 MHz filtering. (c) Windowed output and the scaled window weight. The title reports the band-pass signal-to-noise ratio (SNR). The pulse core is retained unchanged, half-cosine shoulders taper the surrounding samples and the trace is zero outside the outer support. The full-trace path retains the complete noisy waveform; weights given by the inverse quiet-window noise spectral power are introduced only during RF inversion.}
  \label{fig:gate_example}
\end{figure}

Unlike the analytical field reconstruction of Ref.~\cite{refThreePolEField}, which starts from antenna open-circuit voltages, the present reconstruction starts from trigger-selected ADC traces containing the RF-chain response and noise. Figure~\ref{fig:gate_example} illustrates the pulse window and the full-trace treatment. Each processed trace supplies a voltage amplitude for the ADF; the fitted direction is then used in the antenna-response inversion of that trace. The following three treatments compare full-trace noise weighting, suppression of out-of-pulse samples and a noise-free reference:
\begin{itemize}
\item \textit{Full-trace noise-weighted}: the complete noisy ADC trace is supplied to the ADF. During the later RF inversion, a quiet-window spectrum provides frequency- and channel-dependent weights.
\item \textit{L1-gated denoised equal-weight}: the saved station L1 index anchors a two-layer gate before both ADF amplitude extraction and RF inversion. The pulse core from 12 samples before to 24 samples after L1 is unchanged. Half-cosine shoulders extend to 36 samples before and 64 samples after L1, beyond which the trace is zero. Equal inversion weights are used.
\item \textit{Clean equal-weight}: stochastic detector noise is omitted before the ADF and RF inversion, and equal inversion weights are used.
\end{itemize}
The figure labels Full waveform, Denoised and Noiseless denote these three treatments, respectively. The tapered gate preserves samples in its unity region. It suppresses the spectral leakage of a discontinuous hard truncation while retaining a finite noise contribution around the pulse. Its position is fixed by the saved L1 index. The noisy full-trace and trigger-windowed modes share the timing corrections extracted from the complete noisy traces; windowing changes the amplitudes supplied to the ADF and RF inversion while retaining this timing input. The noise-free mode uses its own pulse envelopes. The exact window is given in appendix~\ref{app:gate}.

\subsubsection{Cross-correlation alignment}
\label{sec:timing}

The wavefront reconstruction in Ref.~\cite{refZhang2026E2E} uses pulse-peak times. Our event records instead supply L1 threshold-crossing times, which can identify different parts of the pulse at different stations. We use pairwise waveform cross-correlation to correct these relative offsets before the wavefront fit. Let $\vec r_i$ and $t_i^{\rm L1}$ denote the station position and recorded L1 trigger time. The complete three-channel voltages are filtered to 50--200 MHz and combined into the analytic-signal vector envelope
\begin{equation}
U_i(t)=\left[\sum_{p=1}^{3}|\mathcal H[V_{p,i}](t)|^2\right]^{1/2}.
\label{eq:timing_envelope}
\end{equation}
Here $\mathcal H[V]$ denotes the complex analytic signal formed from the waveform and its Hilbert transform. For each station, a segment extends from 20 samples before to 50 samples after the saved trigger index. We subtract the segment's 15th-percentile baseline and normalise the segment to unit Euclidean norm. Pairwise correlations search lags within $\pm15$ samples; a three-point parabola refines each peak. Pairs with normalised peak correlation $C_{ij}<0.35$ are discarded. The retained lags $\ell_{ij}$ and station offsets $\delta_i$, both in samples, satisfy the weighted least-squares problem below. Its zero-sum constraint fixes the otherwise arbitrary common time offset:
\begin{equation}
\min_{\{\delta_i\}}\sum_{i<j}C_{ij}^2[(\delta_i-\delta_j)-\ell_{ij}]^2,
\qquad\sum_i\delta_i=0.
\label{eq:xcorr_offsets}
\end{equation}
With the 2 ns sampling interval,
\begin{equation}
t_i=t_i^{\rm L1}+(2~\mathrm{ns})\delta_i.
\label{eq:corrected_times}
\end{equation}
Here $N_{\rm ant}$ is the number of participating stations; at least $\max(2,N_{\rm ant}-1)$ usable station pairs are required. The full-trace and trigger-windowed treatments share the timing corrections obtained from noisy traces; the noise-free treatment uses its own envelopes.

\begin{figure}[!htbp]
\centering
\includegraphics[width=\textwidth]{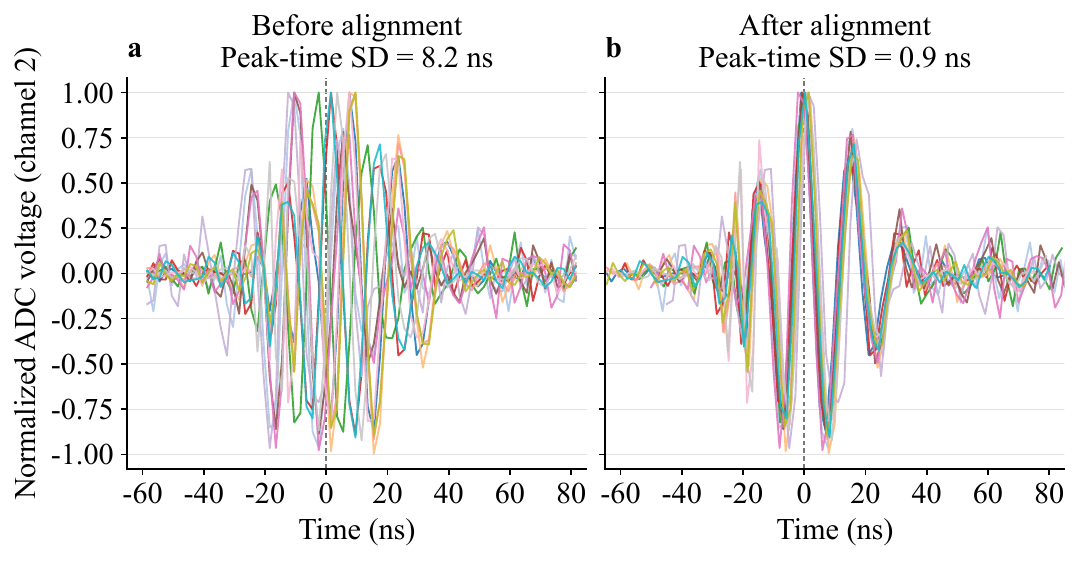}
\caption{Cross-correlation pulse alignment for a representative event. Each curve shows one of the 13 triggered stations. All curves use ADC channel 2, the strongest channel across this event, filtered to 50--200 MHz and normalised by each station's maximum absolute voltage. Panels (a) and (b) show the trigger-relative pulse positions before and after applying the offsets obtained from the three-channel vector envelopes. Both panels use the same common time origin. SD denotes the sample standard deviation of the peak times.}
\label{fig:xcorr_alignment}
\end{figure}

Figure~\ref{fig:xcorr_alignment} illustrates the alignment for an event with 13 triggered stations. The standard deviation of the trigger-relative absolute-peak times decreases from 8.2 ns before alignment to 0.9 ns after applying the cross-correlation offsets.

\subsection{Plane- and spherical-wavefront reconstruction}
\label{sec:wavefront}

We retain the plane- and spherical-wavefront descriptions used in inclined-shower reconstruction~\cite{refDecoene2023,refZhang2026E2E}. The plane-wavefront fit uses the delay gradient across array baselines to provide a seed for the local radio-propagation direction. The spherical-wavefront fit uses the timing curvature associated with a finite source distance to constrain the effective emission point. In place of the array-averaged effective refractive index used in Ref.~\cite{refZhang2026E2E}, we evaluate an optical-path integral for each trial source--station pair to account for refractivity differences between propagation paths. The plane-wave fit minimises
\begin{equation}
\chi^2_{\rm PWF}(\hat k)=\sum_{i<j}\left[\frac{c}{n_0}(t_i-t_j)-\hat k\cdot(\vec r_i-\vec r_j)\right]^2,
\label{eq:pwf}
\end{equation}
where $c$ is the vacuum speed of light, $\hat k$ is the unit downward propagation direction and $n_0=1.0003$ is the constant refractive index used for the plane-wave fit. The spherical-wavefront fit estimates $\vec r_s$ by minimising $\sum_i[t_i-t_0-T_i(\vec r_s)]^2$, with
\begin{equation}
T_i(\vec r_s)=\frac{|\vec r_i-\vec r_s|}{c}\,\bar n_i,
\qquad \bar n_i=\int_0^1 n[h(\vec r_s+u(\vec r_i-\vec r_s))]\,\mathrm du.
\label{eq:path_time}
\end{equation}
Here $T_i$ is the source-to-station travel time, $t_0$ is the common emission-time offset, $u$ parameterises the straight source--station segment, and $h$ returns its altitude; $\bar n_i$ is the path-averaged refractive index. The straight-path optical integral is an approximation to curved-ray travel time. Altitude includes spherical-Earth curvature with radius $6\,370\,949$ m, and $n(h)=1+325\times10^{-6}\exp[-0.1218(h/\mathrm{km})]$. Eight-point Gauss--Legendre quadrature evaluates each path separately. For each trial source, the common intercept is $t_0=\langle t_i-T_i\rangle$. Multiple starting heights of 3, 6, 10, 20, 50 and 80 km are tested; the best solution is refined by bounded nonlinear least squares. The three source coordinates are fitted freely within broad numerical bounds, without a source-distance prior.

The voltage ADF holds the timing-derived source position fixed and fits the particle-axis direction from station voltage amplitudes and source-end ray angles. The resulting source and axis are used in response inversion and footprint mapping.

In the geometry of figure~\ref{fig:pipeline}, $\vec r_{\rm obs}$ is a reference point on the observation plane and $\hat n_{\rm obs}$ its unit normal. With the fitted particle axis $\hat k_{\rm ADF}\equiv\hat v$, the intersection and along-axis distance are
\begin{equation}
\begin{aligned}
\lambda_{\rm obs}&=\frac{(\vec r_{\rm obs}-\vec r_s)\cdot\hat n_{\rm obs}}{\hat k_{\rm ADF}\cdot\hat n_{\rm obs}},\\
\vec r_{\rm pcore}&=\vec r_s+\lambda_{\rm obs}\hat k_{\rm ADF},\qquad d_{\rm rec}=|\lambda_{\rm obs}|.
\end{aligned}
\label{eq:final_core_distance}
\end{equation}
The axis and source define the central-ray construction; $d_{\rm rec}$ enters the event-side amplitude correction. The radio-footprint centre sets the radial origin. Both centres and the source remain fixed during the LDF fit. The timing-quality statistic
\begin{equation}
r_{\rm sph}=\left[\frac{1}{N_{\rm ant}^{\rm used}}\sum_i(t_i-t_0-T_i)^2\right]^{1/2}
\label{eq:rsph}
\end{equation}
is expressed in nanoseconds and used for selection. Here $N_{\rm ant}^{\rm used}$ counts the stations included in this residual sum. The statistic measures the timing residual of the effective-point model.

\subsection{Particle-axis refinement with the voltage angular distribution}

We adapt the ADF framework of Ref.~\cite{refZhang2026E2E} to voltage amplitudes. Whereas that formulation defines its angular coordinates using the straight source-to-station direction, we use the ray tangent at the source to account for the effect of atmospheric refraction on the viewing angle. Fitting in voltage determines the direction required for antenna-response inversion before the electric field is recovered. For each waveform mode, the three-channel voltages are filtered to 50--200 MHz and combined as $V_{{\rm norm},i}(t)=[\sum_pV_{p,i}^2(t)]^{1/2}$. The nonnegative station amplitude is $A_i=\max_t|\mathcal H[V_{{\rm norm},i}](t)|$. This ADF amplitude differs from the vector analytic envelope used for timing in equation~\eqref{eq:timing_envelope}; channel-frequency noise weights enter only the subsequent field inversion.

For each station, a refracted ray connects the fixed effective source to $\vec r_i$ in the spherical atmosphere. Let $\hat u_i^{\rm src}$ be its unit tangent at the source, $\hat v$ the unit trial particle axis, and $\vec B$ the local geomagnetic field. The unit vectors $\hat e_1\parallel\hat v\times\vec B$ and $\hat e_2=\hat v\times\hat e_1$ span the plane normal to the axis. The viewing angle $\omega_i$ and azimuth $\eta_i$ of the source-end tangent in this basis are
\begin{equation}
\begin{aligned}
\omega_i&=\arccos(\hat u_i^{\rm src}\cdot\hat v),\\
\eta_i&=\operatorname{atan2}(\hat u_i^{\rm src}\cdot\hat e_2,\hat u_i^{\rm src}\cdot\hat e_1).
\end{aligned}
\label{eq:ray_adf_angles}
\end{equation}
Geometrical dilution uses the chord distance $L_i=|\vec r_i-\vec r_s|$. The voltage model is
\begin{equation}
\begin{aligned}
A_i^{\rm ADF}={}&\frac{a_0}{L_i}(1+b_{\rm ADF}\sin^2\alpha\cos\eta_i)\\
&\times\left[1+4\left(\frac{(\tan\omega_i/\tan\omega_c)^2-1}{\Delta\omega}\right)^2\right]^{-1}.
\end{aligned}
\label{eq:adf}
\end{equation}
Here $a_0$ is the event amplitude scale, $\alpha$ is the angle between the particle axis and $\vec B$, $\omega_c$ is the ring angle, $\Delta\omega$ is the dimensionless shape-width parameter and $b_{\rm ADF}=0.005$. The width is restricted to $0.03\leq\Delta\omega\leq5$ and the ring angle to $0.20^\circ$--$2.50^\circ$. A Gaussian ring-angle prior with standard deviation $0.35^\circ$ is centred on $\omega_0=\arccos[1/n(\vec r_s)]$. With $N_A$ valid stations and $A_*^2=\langle A_i^2\rangle$, the fit minimises
\begin{equation}
Q_{\rm ADF}=\frac{\sum_i(A_i-A_i^{\rm ADF})^2}{N_AA_*^2}+\left(\frac{\omega_c-\omega_0}{0.35^\circ}\right)^2.
\label{eq:adf_objective}
\end{equation}
The fitted axis determines the particle core and central-ray launch direction. The ADF amplitude scale and ring-shape parameters describe the voltage distribution; energy is subsequently estimated from the recovered electric-field amplitudes.

\subsection{ADC-to-electric-field reconstruction}
\label{sec:rf_inverse}

We retain the analytical three-polarisation inversion of Ref.~\cite{refThreePolEField}, which relates antenna open-circuit voltages to the transverse electric field. To apply it to ADC traces, we first de-embed the loaded antenna/RF network. This step places the signal and the quiet-window noise spectrum at the open-circuit-voltage reference plane required by the inversion. The subsequent direction-dependent antenna inversion uses the geometry supplied by the spherical-wavefront fit and the voltage ADF.

The inverse path summarised in figure~\ref{fig:rf_forward_inverse}(b) acts directly on the stored ADC traces. Each voltage trace is transformed once using a fast Fourier transform (FFT) and processed at its discrete positive-frequency bins. A smooth spectral window retains 50--200 MHz with an 8 MHz transition on those same bins; no interpolation to integer-MHz frequencies is performed. Let $\mathbf T_p(f)$ be the circuit-modelled two-port transfer matrix between the antenna port and ADC for channel $p$~\cite{refNoiseRFI,refGRANDlib}. With $I_{{\rm ADC},p}=V_{{\rm ADC},p}/Z_{{\rm ADC},p}$, the antenna-port state and open-circuit voltage are recovered as
\begin{equation}
\begin{pmatrix}V_{1,p}\\ I_{1,p}\end{pmatrix}
=\mathbf T_p(f)
\begin{pmatrix}V_{{\rm ADC},p}\\ I_{{\rm ADC},p}\end{pmatrix},
\qquad
V_{{\rm oc},p}=I_{1,p}\left(Z_{{\rm in},p}+Z_{{\rm ant},p}\right).
\label{eq:network_inverse}
\end{equation}
Here $V_{1,p}$ and $I_{1,p}$ are the voltage and current at the antenna port, $Z_{{\rm ADC},p}$ is the ADC input impedance, $Z_{{\rm in},p}$ is the network input impedance seen by the antenna, and $Z_{{\rm ant},p}$ is the antenna impedance. The impedances and transfer matrices are complex and evaluated at the discrete FFT frequencies of the corresponding event waveform. Equation~\eqref{eq:network_inverse} de-embeds the same loaded two-port network summarised by $G_p$ in equation~\eqref{eq:forward_chain}. It restores both amplitude and phase at the antenna port. The second relation removes antenna loading to recover $V_{{\rm oc},p}$.

For the three receiving channels, the open-circuit voltages satisfy
\begin{equation}
\mathbf V_{\rm oc}(f)=\mathbf H(f;\theta_{\rm rec},\phi_{\rm rec})
\begin{pmatrix}E_\theta(f)\\E_\phi(f)\end{pmatrix},
\end{equation}
Here $\mathbf V_{\rm oc}$ stacks the three open-circuit voltages. The angles $(\theta_{\rm rec},\phi_{\rm rec})$ specify the direction supplied to the antenna inversion; $E_\theta$ and $E_\phi$ are the two field components transverse to that direction. Row $p$ of the complex $3\times2$ matrix $\mathbf H$ contains the corresponding components of the tabulated vector effective length. The transverse field is reconstructed by
\begin{equation}
\widehat{\mathbf E}_{\theta\phi}=
\left(\mathbf H^\dagger\mathbf W\mathbf H\right)^{-1}
\mathbf H^\dagger\mathbf W\mathbf V_{\rm oc},
\label{eq:efield_wls}
\end{equation}
The dagger denotes conjugate transpose, and $\mathbf W$ is the channel-weight matrix. The normal equations are solved directly at each frequency. If the solver reports a singular matrix, the system is solved using the Moore--Penrose pseudoinverse of the same normal matrix. In the full-trace noise-weighted mode, $\mathbf W$ is diagonal. The pulse-free segment after the pulse is demeaned, transformed to the signal frequency bins and de-embedded to the open-circuit-voltage reference plane. The diagonal weights are
\begin{equation}
\mathbf W(f_k)=\operatorname{diag}_{p=1,2,3}
\left[\frac{1}{|\widetilde V_{{\rm oc},p}^{\rm quiet}(f_k)|^2+\epsilon_{\rm num}}\right],
\label{eq:noise_weights}
\end{equation}
Here $f_k$ is a discrete FFT frequency and $\widetilde V_{{\rm oc},p}^{\rm quiet}$ is the quiet-window voltage spectrum referred to the antenna open circuit; $\epsilon_{\rm num}$ protects against near-zero denominators. This weighting gives less influence to channel-frequency bins with larger quiet-window noise power; the forward response model is unchanged. For the L1-gated denoised and clean traces, $\mathbf W=\mathbf I$. The radial far-field component is set to zero, the transverse solution is rotated into ground coordinates and an inverse real FFT returns $\vec E(t)$.

At each station, the time at which the three-component electric-field norm is maximal is used to sample all three field components. The signed field vector at this time is projected onto the shower-plane basis. Here $\vec v$ denotes the reconstructed particle-propagation direction, equivalent to $\hat v$ above; $x'$ is parallel to $\vec v\times\vec B$ and $y'$ to $\vec v\times(\vec v\times\vec B)$. The LDF observable is
\begin{equation}
E_{x',i}^{\rm pk}=
\left|\vec E_i(t_{\rm vec,pk})\cdot\hat e_{\vec v\times\vec B}\right|,
\qquad
t_{\rm vec,pk}=\underset{t}{\arg\max}\,|\vec E_i(t)|.
\label{eq:vector_peak}
\end{equation}
The same pulse-time definition is used for detector events and reference LDFs. It differs from taking the maximum of the Hilbert envelope of the projected scalar, which is systematically larger for a band-limited pulse. The absolute projection gives a non-negative LDF amplitude while preserving one pulse time for all field components. The $x'$ component is dominated by geomagnetic emission but retains a position-dependent projection of the charge-excess field. We assess its effect through the LDF-fit quality and energy residuals grouped by primary particle and zenith angle.

Whereas radiation-energy estimators integrate the squared field in time and across the footprint~\cite{refAERAEnergy,refRadiationEnergy,refZhang2026E2E}, we fit reference LDFs to the peak amplitudes sampled by the triggered stations. The common pulse time defines one field vector per station, and the fixed reference shapes specify the lateral distribution between the sparse observations. This peak observable also avoids accumulating residual noise and bandpass ringing over a time-integration window.

\subsection{Shower-plane coordinates and reference LDFs}
\label{sec:profiles}

The recovered station amplitudes can be compared with the reference LDFs only after both are expressed in the same shower-plane coordinates and corrected for the early--late effect. We construct a right-handed basis from the reconstructed propagation direction $\vec v$ and the local magnetic field $\vec B$, with $\hat e_{x'}\parallel\vec v\times\vec B$, $\hat e_{y'}\parallel\vec v\times(\vec v\times\vec B)$ and $\hat e_{z'}\parallel\vec v$. A station at ground position $\vec r_i$ is mapped to $\vec q_i$ on the plane normal to $\vec v$ through the radio-footprint centre, along the line connecting the station and the effective emission point $\vec r_s$. The leading early--late asymmetry is corrected using~\cite{refSchlueter2023}
\begin{equation}
g_i=\frac{|\vec r_s-\vec q_i|}{|\vec r_s-\vec r_i|},
\qquad
E_{x',i}^{\rm EL}=\frac{E_{x',i}^{\rm pk}}{g_i}.
\label{eq:early_late}
\end{equation}
The factor $g_i$ is the ratio of the source-to-projected-position and source-to-station distances; the superscript EL denotes the early--late correction. This correction accounts for unequal propagation distances on the two sides of the footprint. The radial distance is $r_{\perp,i}=|\vec q_i-\vec r_{\rm rcore}|$, measured from the radio centre in equation~\eqref{eq:core_shift}. The event and reference LDF use the same amplitude and coordinate definitions.

The energy estimator compares the reconstructed amplitudes with ZHAireS-based reference LDFs constructed in the FARSim coordinate framework~\cite{refZHAireS,refFARSim}. Each simulated three-component field trace is filtered to 50--200 MHz directly at its discrete FFT frequencies. The signed vector is sampled at the maximum of $|\vec E(t)|$ and projected onto $\hat e_{x'}$, using the same observable as equation~\eqref{eq:vector_peak}. Observer coordinates and amplitudes are then corrected for the early--late effect according to equation~\eqref{eq:early_late}, and the radial origin is shifted according to equation~\eqref{eq:core_shift}.

Let $\phi'$ be the shower-plane azimuth measured from $x'$. The leading geomagnetic and radially polarised charge-excess field amplitudes, $E_{\rm geo}$ and $E_{\rm ce}$, contribute to the total field (superscript tot) as~\cite{refAugerPolarization,refLOFARPolarization,refSchlueter2023}
\begin{equation}
\begin{aligned}
E_{x'}^{\rm tot}(r_\perp,\phi')&=E_{\rm geo}(r_\perp)+E_{\rm ce}(r_\perp)\cos\phi',\\
E_{y'}^{\rm tot}(r_\perp,\phi')&=E_{\rm ce}(r_\perp)\sin\phi'.
\end{aligned}
\label{eq:farsim_component_decomposition}
\end{equation}
The four symmetric star-shaped arms at $\phi'=0,\pi/2,\pi,3\pi/2$ provide the azimuthal average used below and satisfy $\langle\cos\phi'\rangle_{\phi'}=0$. Here $s$ indexes the ten independent shower seeds, and the superscript $\mathrm{tot,EL}$ denotes the total $x'$ component after the early--late correction. The reference LDF for primary hypothesis $k\in\{\mathrm{Fe},\mathrm p\}$ and zenith node $\theta_j$ is therefore defined by
\begin{equation}
M^k(r_\perp;\theta_j)=
\frac{1}{N_s}\sum_{s=1}^{N_s}
\left\langle
E_{x',s}^{\mathrm{tot,EL}}(r_\perp,\phi';\theta_j,k)
\right\rangle_{\phi'},
\qquad N_s=10,
\label{eq:farsim_radial_profile}
\end{equation}
which cancels the leading charge-excess harmonic. Averaging corresponding observers over ten independent shower seeds suppresses shower-to-shower fluctuations without smoothing the Cherenkov-ring structure. Each profile also stores the simulated source distance, atmospheric density and geomagnetic direction needed by the 50--200 MHz peak-amplitude correction in section~\ref{sec:coherence}.  Because a detector event samples only a finite and generally asymmetric set of azimuths, its one-dimensional $x'$ distribution can retain a charge-excess contribution. 

The iron and proton reference families are tabulated at integer zenith angles, a reference azimuth of $45^\circ$, a reference energy $E_0=0.316$ EeV and 176 transformed observer positions. The principal detector analysis uses nodes no greater than $85^\circ$. The stored $88^\circ$--$89^\circ$ extension comes from a separate simulation batch and is not used by this detector-level result. For each primary hypothesis, the nearest integer node to the reconstructed zenith, restricted to $45^\circ$--$85^\circ$, is selected for detector-level reconstruction and the reference LDF is linearly interpolated at each station:
\begin{equation}
M_i^k=M^k(r_{\perp,i};\theta_{\rm node})
\frac{\sin\alpha_{\rm event}}{\sin\alpha_{\rm ref}}.
\label{eq:station_template}
\end{equation}
Here $\alpha_{\rm event}$ and $\alpha_{\rm ref}$ are the event and reference geomagnetic angles. Equation~\eqref{eq:station_template} accounts for the leading geomagnetic scaling~\cite{refAERAEnergy,refSchlueter2023}. Section~\ref{sec:coherence} describes the subsequent correction for differences in source distance, atmospheric density and geomagnetic deflection between an event and its reference LDF.

\subsection{LDF normalisation and energy estimator}
\label{sec:energy}

For stations within the radial range covered by both the iron and proton profiles, let $E_i\equiv |E_{x',i}^{\rm EL}|$ and let $M_i^k$ denote the interpolated profile from equation~\eqref{eq:station_template}. The corrected amplitudes are now compared with the reference values at the same station coordinates. Geometry, station selection and LDF shape are fixed before the energy fit. For each primary hypothesis $k$, the only free parameter is the dimensionless LDF normalisation
\begin{equation}
\begin{aligned}
\hat a_k&=\frac{\sum_i w_i E_iM_i^k}{\sum_i w_i(M_i^k)^2},\\
w_i&=\operatorname{clip}\!\left[\left(\frac{E_i}{\operatorname{median}_j E_j}\right)^2,0.25,4\right].
\end{aligned}
\label{eq:profile_scale}
\end{equation}
The operator $\operatorname{clip}[x,a,b]$ bounds $x$ to the interval $[a,b]$; the median is taken over the selected stations. These weights emphasise stations with larger recovered pulse amplitudes while limiting the leverage of individual stations. They are distinct from the channel-frequency noise weights in RF inversion. No extra template-to-event $1/d$ amplitude multiplier is used after early--late mapping. The source position, particle core and station mask remain fixed during this zero-intercept weighted least-squares fit. Electric-field simulations show that the common-time peak amplitude scales with primary energy to the power $\gamma=1.035$ over the simulated range. If $\widetilde H_k^{(m)}$ is the event-to-reference peak-amplitude correction defined in section~\ref{sec:coherence}, where $m={\rm cal}$ denotes the empirical correction used in the detector analysis and $m={\rm mix}$ denotes the semi-analytic Fresnel approximation evaluated only in the field-level test, the LDF-based energy estimate is
\begin{equation}
E_k^{(m)}=E_0\left(\frac{\hat a_k}{\widetilde H_k^{(m)}}\right)^{1/\gamma},
\qquad E_0=0.316~{\rm EeV},\quad\gamma=1.035.
\label{eq:profile_energy}
\end{equation}
The final estimate is the fixed equal-weight combination
\begin{equation}
E_{\rm rec}^{(m)}=\frac{1}{2}\left(E_{\rm Fe}^{(m)}+E_{\rm p}^{(m)}\right).
\label{eq:energy_blend}
\end{equation}
Unless stated otherwise, $E_{\rm rec}$ uses the empirical peak-amplitude correction defined below. Because the primary mass is unknown, equation~\eqref{eq:energy_blend} averages the proton and iron energy estimates with fixed equal weights. This avoids selecting a reference from fit residuals that depend on noise. The same estimator is used for every event and waveform treatment.

\subsection{Event-to-reference correction of the 50--200 MHz peak amplitude}
\label{sec:coherence}

The $\sin\alpha$ factor accounts for the leading geomagnetic scaling. The finite-band peak amplitude also depends on the coherence of the radiation. Magnetic deflection, the spatial extent of the emitting particles and propagation-time differences control the coherence of the superposed fields~\cite{refChiche2024}. Showers of equal energy can therefore have different 50--200 MHz peak amplitudes when their source distance, atmospheric density or geomagnetic angle differs. Previous density-dependent corrections act on geomagnetic radiation energy~\cite{refZhang2026E2E}. Here we calibrate the event-to-reference peak-amplitude ratio in the fixed 50--200 MHz band after normalising both distributions to the same shower energy. Evaluating the correction for both the event and its reference accounts for the amplitude dependence already contained in the stored LDF. The empirical function is used in the energy estimator; the semi-analytic Fresnel model tests whether finite-source interference can explain the calibrated dependence.

\subsubsection{Empirical peak-amplitude correction}

Following the analytical coherence-scale estimate in Ref.~\cite{refChiche2024}, and fixing the characteristic electron energy to the critical energy $\epsilon_e=88$ MeV, we define the dimensionless variable $R$ as the squared ratio of the magnetically broadened shower scale $l_{\rm lat}$ to the spatial coherence length $l_{\rm coh}$:
\begin{equation}
R \equiv \left(\frac{l_{\rm lat}}{l_{\rm coh}}\right)^2
=
0.018
\left(\frac{\nu_{\rm eff}}{50~{\rm MHz}}\right)
\left(\frac{B\sin\alpha}{50~\mu{\rm T}}\right)^2
\left(\frac{\rho_s}{1~{\rm kg\,m^{-3}}}\right)^{-4}
\left(\frac{d_{\rm obs}}{10~{\rm km}}\right)^{-1},
\label{eq:coherence_variable}
\end{equation}
where $l_{\rm coh}=\sqrt{(c/\nu)d_{\rm obs}}$, $\nu_{\rm eff}=99$ MHz represents the fixed 50--200 MHz band, $B=55.348~\mu$T and $d_{\rm obs}$ is the distance from the effective emission point to the observation plane. Here $\rho_s$ is the atmospheric density at that fitted point. The analytical scale estimate uses this density to characterise the effective emission region. The magnetic-deflection estimate gives $l_{\rm lat}\propto B\sin\alpha\,\epsilon_e^{-1}\rho_s^{-2}$, which yields the $\sin^2\alpha$, $\rho_s^{-4}$ and $d_{\rm obs}^{-1}$ dependences in equation~\eqref{eq:coherence_variable}~\cite{refChiche2024}. Replacing the monochromatic frequency by $\nu_{\rm eff}$ and using $R$ to parameterise a bandwidth-limited peak amplitude are specific to this work. We parameterise the dependence of the finite-band peak amplitude on $R$ as
\begin{equation}
C(R;R_0,p)=
\left[1+\left(\frac{R}{R_0}\right)^p\right]^{-1/(2p)}.
\label{eq:coherence_response}
\end{equation}
Here $R_0$ is a dimensionless transition scale and $p$ controls the shape of the transition. The function $C$ decreases with increasing $R$. The energy estimator uses the event-to-reference ratio defined below. For a detector event, $\alpha_{\rm event}$ is calculated from the reconstructed direction, $d_{\rm event}\equiv d_{\rm rec}$ is the source distance returned by equation~\eqref{eq:final_core_distance}, and $\rho_{\rm event}$ is evaluated at the fitted emission point with the fixed atmospheric profile. Thus
\begin{equation}
C_{\rm event}=
C\!\left[
R(\alpha_{\rm event},\rho_{\rm event},d_{\rm event});R_0,p
\right]
\label{eq:event_response}
\end{equation}
is determined entirely from reconstructed event quantities and fixed site information. The corresponding value for reference family $k$ uses the simulated source geometry of each of its ten seeds,
\begin{equation}
\overline C_{{\rm ref},k}=
\frac{1}{10}\sum_{s=1}^{10}
C\!\left[
R(\alpha_{\rm ref}(\theta_{\rm node},\phi_{\rm node}),
\rho_{k,s},d_{k,s});R_0,p
\right].
\label{eq:reference_response}
\end{equation}
Here $\theta_{\rm node}$ and $\phi_{\rm node}$ are the zenith and azimuth of the reference-library node, respectively; the present library uses the fixed value $\phi_{\rm node}=45^\circ$. The angle $\alpha_{\rm ref}$ is the resulting angle between the reference shower axis and the site magnetic field. The quantities $\rho_{k,s}$ and $d_{k,s}$ are the atmospheric density and source-to-observation-plane distance stored for seed $s$ of reference family $k$.
The event and reference quantities remain distinct because each stored profile already contains the finite-band peak-amplitude response associated with its own simulated geometry. The event-to-reference correction factor is therefore
\begin{equation}
H_k^{\rm cal} =
\mathcal{A}
\frac{C_{\rm event}}
{\overline C_{{\rm ref},k}}.
\label{eq:relative_response_cal}
\end{equation}
The event-to-reference factor $H_k^{\rm cal}$ uses the normalisation $\mathcal A$, determined from the electric-field calibration sample and held fixed during validation.

Ref.~\cite{refChiche2024} identifies $\theta\gtrsim65^\circ$ as the lower boundary of the very-inclined regime in which coherence-related changes become relevant for a geomagnetic field of order $50~\mu$T. We adopt $65^\circ$ as the operational lower boundary for applying the event-to-reference correction: the uncorrected LDF normalisation is used below this angle and the correction is applied at and above it,
\begin{equation}
\widetilde H_k^{(m)}=
\begin{cases}
1, & \theta_{\rm rec}<65^\circ,\\
H_k^{(m)}, & \theta_{\rm rec}\geq65^\circ,
\end{cases}
\label{eq:coherence_gate}
\end{equation}
for both the empirical ($m={\rm cal}$) and semi-analytic ($m={\rm mix}$) functions. The $65^\circ$ threshold specifies the angular range in which the correction is applied. Within that range, the continuous function $C(R)$ describes its dependence on frequency, $B\sin\alpha$, atmospheric density and source distance. Detector events use the zenith reconstructed independently in each waveform mode; the field-level test below uses the simulated input zenith because the complete simulated input geometry is supplied directly.

The parameters in equation~\eqref{eq:coherence_response} were estimated from all 2,500 electric-field calibration showers. For each shower $j$, the measured event-to-reference amplitude ratio $H_j^{\rm obs}$ was obtained by fitting the energy-normalised event LDF to the reference LDF for the simulated primary over their common radial support. Parameter estimation minimises a robust loss applied to the logarithmic residual $\Delta_j=\ln[H_j^{\rm obs}/H_{k(j)}^{\rm cal}]$, with $u_j=\Delta_j/0.05$ and equal weight for every shower; $k(j)$ denotes shower $j$'s simulated primary type. The loss is approximately quadratic for $|u_j|\ll1$ and approximately linear for large $|u_j|$, reducing the influence of isolated large residuals. The numerical scale 0.05 sets this transition. The fitted parameters were fixed before validation:
\begin{equation}
\mathcal A=0.996725,\qquad
R_0=3.012370,\qquad
p=0.701326 .
\label{eq:coherence_parameters}
\end{equation}
The same correction parameters are used for proton and iron showers. The reference family determines the reference LDF and the reference-side value in equation~\eqref{eq:reference_response}. The fitted normalisation is close to unity for the amplitude--energy scaling defined in equation~\eqref{eq:profile_energy}.

\subsubsection{Semi-analytic Fresnel interpretation of the empirical correction}

To examine whether finite-source interference can account for the empirical correction, we construct a semi-analytic Fresnel approximation based on the finite-track superposition used in microscopic radio-emission calculations. For one projected polarisation, the complex field can be written schematically as
\begin{equation}
\widetilde E(\omega,\mathbf r_o)=
\int \mathrm d\boldsymbol\xi\,
J(\boldsymbol\xi,\omega)
\exp\left[i\omega\Delta t(\boldsymbol\xi,\mathbf r_o)\right],
\label{eq:track_field}
\end{equation}
Here $\omega$ is angular frequency, not the ADF viewing angle $\omega_i$, and $\mathbf r_o$ is the observer position. The variable $\boldsymbol\xi$ describes the emitting phase-space coordinates, $J$ contains the finite-track amplitude and polarisation factors, and $\Delta t$ includes emission time and atmospheric optical-path delay. Equation~\eqref{eq:track_field} approximates the finite-track amplitude-and-phase superposition underlying ZHAireS and CoREAS~\cite{refZHAireS,refCoREAS}. Near a stationary Cherenkov region, let the effective source distribution be Gaussian with covariance $\boldsymbol\Sigma$ and expand the delay to second order,
\begin{equation}
\Delta t=\Delta t_0+\mathbf g^{\mathsf T}\boldsymbol\xi
+\frac{1}{2}\boldsymbol\xi^{\mathsf T}\mathbf Q\boldsymbol\xi.
\label{eq:quadratic_delay}
\end{equation}
Here $\Delta t_0$ is the delay at the expansion point, $\mathbf g$ is its gradient with respect to $\boldsymbol\xi$, and $\mathbf Q$ is the Hessian of that delay. The normalised Gaussian integral is then
\begin{equation}
\mathcal H(\omega)=
\det\!\left(\mathbf I-i\omega\boldsymbol\Sigma\mathbf Q\right)^{-1/2}
\exp\!\left[-\frac{\omega^2}{2}\mathbf g^{\mathsf T}
\left(\boldsymbol\Sigma^{-1}-i\omega\mathbf Q\right)^{-1}\mathbf g\right].
\label{eq:gaussian_fresnel_kernel}
\end{equation}
Here $\mathbf I$ is the identity matrix, the superscript $\mathsf T$ denotes transpose, and $\det$ denotes a determinant. The complex factor $\mathcal H(\omega)$ describes the amplitude and phase changes caused by interference within the finite source; it is distinct from the analytic-signal operator $\mathcal H[V]$ used for waveform envelopes. Atmospheric refractivity determines the stationary point and the curvature of the propagation delay, linking these changes to source distance and density~\cite{refWerner2012}.

The one-dimensional stationary limit of equation~\eqref{eq:gaussian_fresnel_kernel} motivates the kernel
\begin{equation}
h(R;s)=\left(1-iR/s\right)^{-1/2},
\label{eq:positive_kernel}
\end{equation}
where $R$ is the dimensionless variable defined in equation~\eqref{eq:coherence_variable} and $s>0$ is an effective Fresnel scale. We represent the range of particle energies, transverse displacements and emission heights by assuming a lognormal distribution of positive effective source scales and averaging the kernel over it,
\begin{align}
s(z)&=R_c\exp(\sigma_s z),\qquad z\sim\mathcal N(0,1),
\label{eq:positive_scales}\\
\mathcal H_{\rm mix}(R)&=
\mathbb E_z\!\left[h\!\left(R;s(z)\right)\right],
\qquad
C_{\rm mix}(R)=\left|\mathcal H_{\rm mix}(R)\right|.
\label{eq:positive_mixture}
\end{align}
Equations~\eqref{eq:positive_kernel}--\eqref{eq:positive_mixture} define a phenomenological semi-analytic approximation to the finite-band peak-amplitude dependence.  The restriction $s>0$ keeps the effective source scales positive; $R_c$ is the median scale and $\sigma_s$ is the standard deviation of its logarithm; $\mathbb E_z$ averages over the standard normal variable $z$. Here $s$ denotes a continuous source scale, rather than the discrete reference-shower index used earlier.

The semi-analytic parameters are fitted to the same complete 2,500-shower calibration sample after equations~\eqref{eq:positive_kernel}--\eqref{eq:positive_mixture} are fixed. An overall amplitude was allowed only to check the normalisation and converged to 0.99960; it is fixed to one in all reported evaluations. The energy-amplitude exponent is fixed independently to $\gamma=1.035$. The final shape parameters are
\begin{equation}
R_c=1.65808,\qquad \sigma_s=1.68839,\qquad \mathcal A_{\rm mix}\equiv1.
\label{eq:positive_parameters}
\end{equation}

For reference family $k$, the event-to-reference correction factor is
\begin{equation}
H_k^{\rm mix}=
\frac{C_{\rm mix}(R_{\rm event})}
{10^{-1}\sum_{s=1}^{10}C_{\rm mix}(R_{{\rm ref},k,s})}.
\label{eq:relative_response_mix}
\end{equation}
The semi-analytic function is evaluated here only in the field-level test, where $R_{\rm event}$ uses the simulation-input geometry and the denominator retains each reference seed's fixed simulated geometry. The function $C_{\rm mix}$ supplies the event and reference responses in equation~\eqref{eq:relative_response_mix}; their ratio enters the energy conversion in equation~\eqref{eq:profile_energy}.

\subsection{Event selection and performance metrics}

For profile-shape selection, the proton and iron reference distributions are combined with equal weights after applying their event-to-reference corrections:
\begin{equation}
M_i^{\rm eq}=\tfrac12(H_{\rm Fe}^{\rm cal}M_i^{\rm Fe}+H_{\rm p}^{\rm cal}M_i^{\rm p}).
\end{equation}
Its fitted scale $\hat a_{\rm eq}$ is obtained by substituting $M_i^{\rm eq}$ for $M_i^k$ in equation~\eqref{eq:profile_scale}, with the same station weights. With $\bar E_w=\sum_iw_iE_i/\sum_iw_i$, the weighted coefficient of determination is
\begin{equation}
R_{\rm eq}^2=1-\frac{\sum_iw_i(E_i-\hat a_{\rm eq}M_i^{\rm eq})^2}{\sum_iw_i(E_i-\bar E_w)^2}.
\label{eq:r2}
\end{equation}
Each mode must yield finite geometry and energy, at least five stations within the common reference-profile support, $r_{\rm sph}<30$ ns and $R_{\rm eq}^2\geq0.75$. The same quality threshold is applied to all three modes. The comparison retains their event-wise intersection; station masks may differ between modes but are fixed before evaluating the energy residual.

The principal performance interval is
\begin{equation}
60^\circ\leq\theta_{\rm sim}<85^\circ.
\label{eq:primary_domain}
\end{equation}
The simulated input zenith defines the reporting interval. Let
\begin{equation}
\delta_E=(E_{\rm rec}-E_{\rm sim})/E_{\rm sim}.
\end{equation}
Here $E_{\rm sim}$ is the simulated primary energy. We report the residual mean, sample standard deviation and root-mean-square error (RMSE), $\mathrm{RMSE}=\sqrt{\langle\delta_E^2\rangle}$. Directional error $\Delta\Psi$ is the great-circle separation of the simulated and reconstructed directions. Energy-residual error bars in zenith bins denote standard errors of the mean; the direction plot instead shows the median and 68\% quantile as specified in its caption.

\section{Results}
\label{sec:results}

\subsection{Field-level performance limit with simulation-input geometry}
\label{sec:field_benchmark}

With simulated electric fields and input shower geometry, the field-level test assesses the energy performance attainable with the chosen reference LDFs and amplitude corrections. The input excludes detector response, noise and geometry-reconstruction errors, and the simulated primary label selects the reference LDF. This ideal-input benchmark precedes the evaluation of the complete ADC reconstruction.

\begin{figure}[!htbp]
  \centering
  \includegraphics[width=0.94\textwidth]{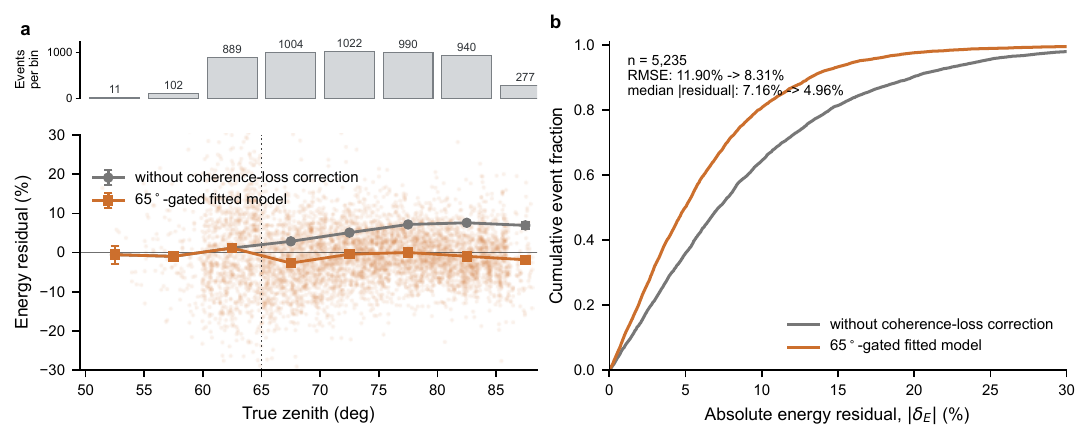}
  \caption{Field-level test of the empirical 50--200 MHz peak-amplitude correction using simulated electric fields and input geometry, with the application threshold in equation~\eqref{eq:coherence_gate}. (a) Bars in the upper strip show the number of events in each $5^\circ$ input-zenith bin. Translucent points in the lower panel are individual post-correction residuals; curves compare bin means before and after correction, with standard errors of the mean. The dotted vertical line marks $65^\circ$. (b) Fraction of events with an absolute energy residual no greater than the value on the horizontal axis. The 5,235 events satisfy $\theta\leq88^\circ$ and $N_{\rm fit}\geq5$. The simulated primary label selects the corresponding reference LDF only in this test. Proton and iron share the parameters in equation~\eqref{eq:coherence_parameters}.}
  \label{fig:empirical_ceiling}
\end{figure}

Figure~\ref{fig:empirical_ceiling} shows the field-level test of the empirical correction on the independent 6,000-shower fixed-array sample. Requiring $\theta\leq88^\circ$ and at least five valid observers leaves 5,235 events. Relative to the uncorrected reconstruction, equation~\eqref{eq:coherence_gate} changes the mean residual from $+4.83\%$ to $-0.67\%$, reduces the energy RMSE from 11.90\% to 8.31\% and the median absolute residual from 7.16\% to 4.96\%, and reduces the zenith slope from $0.322\%$ to $-0.020\%$ per degree. Applying the empirical function at all zenith angles gives a slightly smaller RMSE of 8.19\% but a larger residual zenith slope of $+0.137\%$ per degree. Thus the $65^\circ$ application threshold trades a 0.12-percentage-point increase in RMSE for a mean closer to zero and a substantially flatter zenith dependence. 

\begin{figure}[!htbp]
  \centering
  \includegraphics[width=0.94\textwidth]{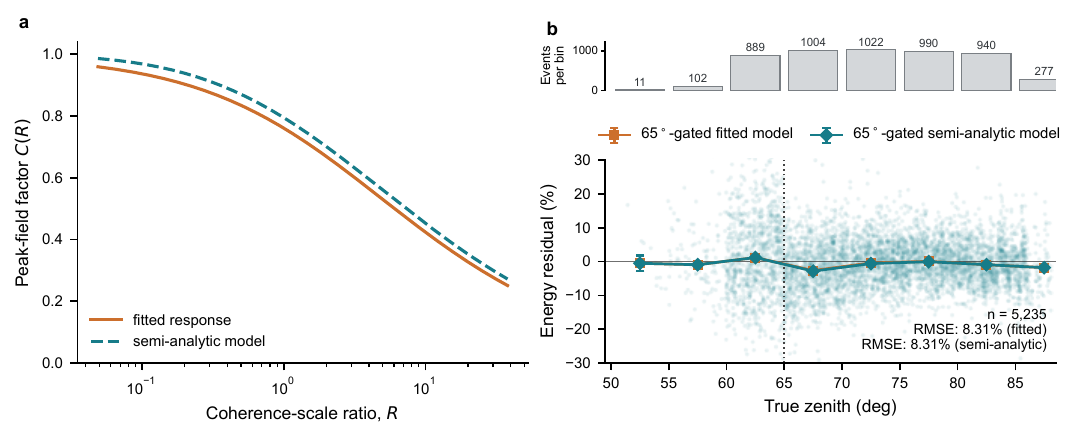}
  \caption{Comparison of the empirical and semi-analytic 50--200 MHz peak-amplitude corrections under equation~\eqref{eq:coherence_gate}. (a) The two functions $C(R)$ as a function of the dimensionless variable $R$; the functions themselves are continuous and are not altered by the $65^\circ$ application threshold. (b) Bars in the upper strip show the number of events in each $5^\circ$ input-zenith bin. Translucent points in the lower panel are individual semi-analytic residuals; curves compare empirical and semi-analytic bin means in the same 5,235-event field-level test using simulated input geometry, with standard errors of the mean. The dotted line marks $65^\circ$. Both calculations select the corresponding primary reference LDF but use one parameter set per function for both primaries. The semi-analytic normalisation is fixed to unity.}
  \label{fig:physical_ceiling}
\end{figure}

Figure~\ref{fig:physical_ceiling} compares the empirical and semi-analytic corrections. In the same 5,235-event field-level test using the simulated input geometry, the semi-analytic correction with the $65^\circ$ application threshold gives a mean residual of $-0.76\%$, an RMSE of 8.31\%, a median absolute residual of 4.95\% and a zenith slope of $-0.018\%$ per degree. The corresponding empirical-function values are $-0.67\%$, 8.31\%, 4.96\% and $-0.020\%$ per degree. The agreement indicates that the Fresnel model with a lognormal distribution of source scales reproduces the energy-correction trend of the empirical function.

\subsection{Common detector-level event sample}

Of the 4,384 triggered showers, 4,383 have finite results in all three waveform modes. The common quality requirements retain 3,119 full-trace, 3,343 trigger-windowed and 3,576 noise-free results. Their intersection contains 3,083 events, including 3,073 with $60^\circ\leq\theta_{\rm sim}<85^\circ$. All detector-level direction and energy comparisons use this common sample (table~\ref{tab:cutflow}).

\begin{table}[!htbp]
\centering
\caption{Event counts with $N_{\rm fit}\geq5$, $r_{\rm sph}<30$ ns and weighted $R_{\rm eq}^2\geq0.75$ in every mode.}
\label{tab:cutflow}
\setlength{\tabcolsep}{7pt}
\renewcommand{\arraystretch}{1.15}
\begin{tabular}{lr}
\toprule
Stage & $N$\\
\midrule
Triggered input & 4384\\
Complete three-mode records & 4383\\
Full-trace quality & 3119\\
Trigger-window quality & 3343\\
Noise-free quality & 3576\\
Three-mode intersection & 3083\\
Common $60^\circ\leq\theta_{\rm sim}<85^\circ$ & 3073\\
\bottomrule
\end{tabular}
\end{table}

\subsection{Arrival direction}

\begin{figure}[!htbp]\centering
\includegraphics[width=\textwidth]{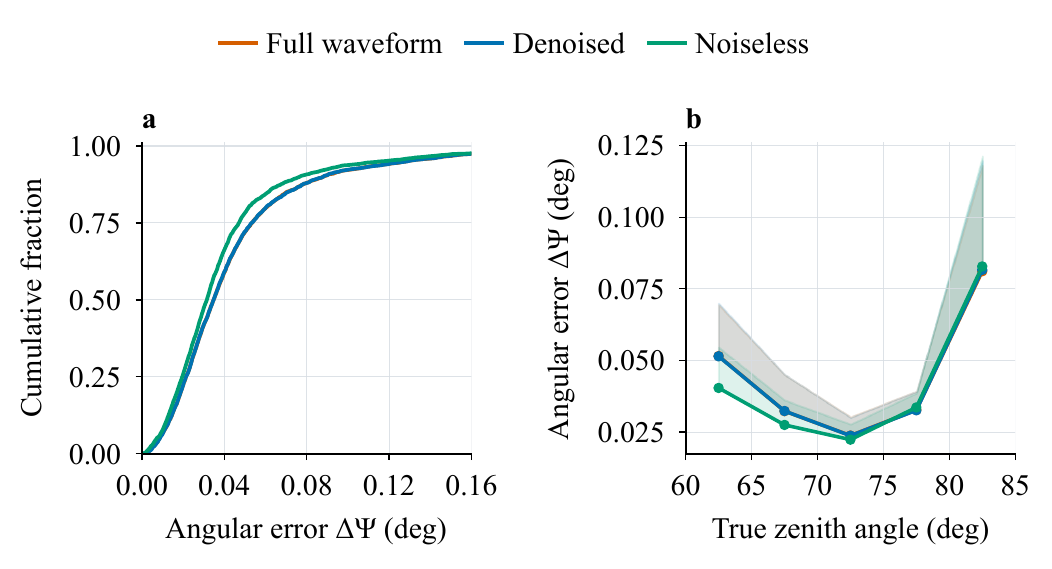}
\caption{Direction reconstruction on the common 3,073-event sample. (a) Cumulative distribution of the opening-angle error. (b) Median opening-angle error in input-zenith bins; shaded regions extend to the 68\% quantile. All three modes use the same event set.}
\label{fig:direction}\end{figure}

Table~\ref{tab:direction_metrics} and figure~\ref{fig:direction} give the directional residuals. The median opening angles are $0.0349^\circ$, $0.0348^\circ$ and $0.0315^\circ$ for full-trace, trigger-windowed and noise-free inputs. Their 68\% quantiles are $0.0464^\circ$, $0.0464^\circ$ and $0.0412^\circ$. All three zenith residual means are about $-0.023^\circ$. The two noisy modes share the cross-correlation timing corrections described in section~\ref{sec:timing}.

\begin{table}[!htbp]
\centering
\caption{Directional residuals for the common 3,073 events, in degrees. $\Delta\theta=\theta_{\rm rec}-\theta_{\rm sim}$; SD denotes the sample standard deviation; P68 and P95 denote the 68th and 95th percentiles of the opening angle.}
\label{tab:direction_metrics}
\setlength{\tabcolsep}{7pt}
\renewcommand{\arraystretch}{1.15}
\begin{tabular}{lrrrrr}
\toprule
Waveform & Mean $\Delta\theta$ & SD $\Delta\theta$ & Median $\Delta\Psi$ & P68 & P95\\
\midrule
Full trace & -0.0234 & 0.0452 & 0.0349 & 0.0464 & 0.1285\\
Trigger window & -0.0233 & 0.0450 & 0.0348 & 0.0464 & 0.1287\\
Noise free & -0.0237 & 0.0418 & 0.0315 & 0.0412 & 0.1160\\
\bottomrule
\end{tabular}
\end{table}

\subsection{Energy reconstruction}

\begin{figure}[!htbp]\centering
\includegraphics[width=\textwidth]{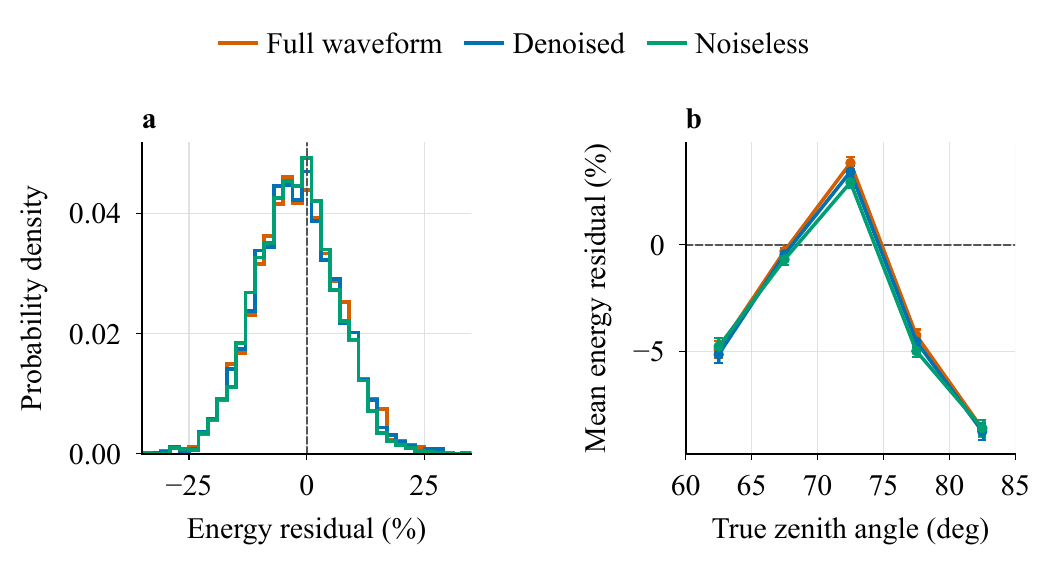}
\caption{Detector-level energy reconstruction on the common sample. (a) Energy residual distributions for the three waveform modes. (b) Mean residual in input-zenith bins. The energy estimator combines the proton and iron energy estimates with fixed equal weights.}
\label{fig:energy}\end{figure}

Figure~\ref{fig:energy} shows the residual distributions and their input-zenith dependence; table~\ref{tab:energy_metrics} gives the summary statistics. For the full-trace, trigger-windowed and noise-free treatments, the mean energy residuals are $-1.86\%$, $-2.12\%$ and $-2.30\%$, with sample standard deviations of 8.96\%, 8.83\% and 8.46\%, respectively. The corresponding RMSE values are 9.15\%, 9.08\% and 8.77\%. These detector-level results use reconstructed geometry; the field-level benchmark in section~\ref{sec:field_benchmark} uses simulation-input geometry and its own selection.

\begin{table}[!htbp]
\centering
\caption{Energy residuals for the common 3,073 events. All values are percentages. The standard deviation uses $N-1$. Mode names are abbreviated as full trace, trigger window and noise free.}
\label{tab:energy_metrics}
\setlength{\tabcolsep}{7pt}
\renewcommand{\arraystretch}{1.15}
\begin{tabular}{lrrr}
\toprule
Waveform & Mean & SD & RMSE\\
\midrule
Full trace & -1.86 & 8.96 & 9.15\\
Trigger window & -2.12 & 8.83 & 9.08\\
Noise free & -2.30 & 8.46 & 8.77\\
\bottomrule
\end{tabular}
\end{table}

\subsection{Effect of the peak-amplitude correction}

\begin{figure}[!htbp]\centering
\includegraphics[width=\textwidth]{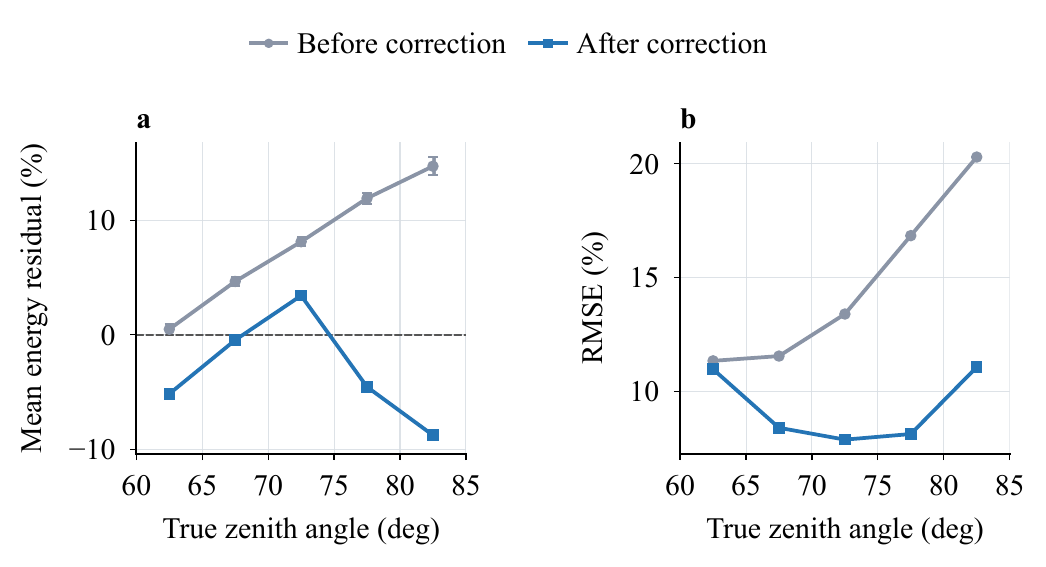}
\caption{Paired effect of the empirical peak-amplitude correction on the common sample, shown for the trigger-windowed treatment. (a) Mean energy residual and (b) RMSE in input-zenith bins, before and after correction. The same selected events and fitted LDF scales enter both calculations.}
\label{fig:coherence_detector}\end{figure}

Figure~\ref{fig:coherence_detector} compares the trigger-windowed energy residuals before and after correction. The paired comparison removes only $H_k^{\rm cal}$ from the energy conversion, retaining the selected events, geometry, fitted scales and station masks. Uncorrected RMSE values are 14.42\%, 14.20\% and 13.94\% for full-trace, trigger-windowed and noise-free inputs, respectively. The empirical correction reduces them to 9.15\%, 9.08\% and 8.77\% (table~\ref{tab:coherence_detector}). 

\begin{table}[!htbp]
\centering
\caption{Paired amplitude-correction comparison on the fixed 3,073-event sample. Values are percentages; selection and fitted scales are held fixed.}
\label{tab:coherence_detector}
\setlength{\tabcolsep}{7pt}
\renewcommand{\arraystretch}{1.15}
\begin{tabular}{llrrr}
\toprule
Waveform & Correction & Mean & SD & RMSE\\
\midrule
Full trace & Without & +7.41 & 12.37 & 14.42\\
Full trace & With & -1.86 & 8.96 & 9.15\\
Trigger window & Without & +7.14 & 12.27 & 14.20\\
Trigger window & With & -2.12 & 8.83 & 9.08\\
Noise free & Without & +6.93 & 12.10 & 13.94\\
Noise free & With & -2.30 & 8.46 & 8.77\\
\bottomrule
\end{tabular}
\end{table}

\subsection{Dependence on zenith angle and primary particle}

As shown in figure~\ref{fig:coherence_detector}(a), the corrected mean energy residual still depends on zenith angle. For the trigger-windowed treatment, the mean residual is $-5.14\%$, $-0.45\%$, $+3.42\%$, $-4.54\%$ and $-8.75\%$ in successive five-degree bins from $60^\circ$ to $85^\circ$. Reconstructed geometry and sparse spatial sampling are possible contributors to the high-inclination negative residual.

The sample contains 1,474 iron and 1,599 proton showers. In the trigger-windowed mode, matching the reference to the simulated primary gives means of $-1.36\%$ for iron and $-3.15\%$ for protons. Using the other reference gives mean residuals of $-7.00\%$ and $+2.73\%$, respectively. Fixed equal weighting gives $-4.18\%$ for iron and $-0.21\%$ for protons (table~\ref{tab:primary_metrics}). The different group means show the primary-particle dependence of the energy residuals.

\begin{table}[!htbp]
\centering
\caption{Primary-group diagnostics in the trigger-windowed mode. Labels are used only for grouping; energy residuals are percentages.}
\label{tab:primary_metrics}
\setlength{\tabcolsep}{7pt}
\renewcommand{\arraystretch}{1.15}
\begin{tabular}{llrrrr}
\toprule
Primary & Reference & $N$ & Mean & SD & RMSE\\
\midrule
Iron & Iron reference & 1474 & -1.36 & 8.38 & 8.49\\
Iron & Proton reference & 1474 & -7.00 & 8.36 & 10.90\\
Iron & Fixed half & 1474 & -4.18 & 8.16 & 9.16\\
Proton & Iron reference & 1599 & +2.73 & 9.55 & 9.93\\
Proton & Proton reference & 1599 & -3.15 & 8.80 & 9.35\\
Proton & Fixed half & 1599 & -0.21 & 9.00 & 9.00\\
\bottomrule
\end{tabular}
\end{table}

\section{Discussion}
\label{sec:discussion}

The common-event comparison shows that waveform treatment mainly changes the spread of the reconstructed energies: the standard deviation rises from 8.46\% without noise to 8.96\% with full-trace noise weighting and falls to 8.83\% with trigger-window processing. The matched-response control in appendix~\ref{app:audits} shows that numerical RF inversion adds little mean energy bias when the simulation-input geometry is supplied. 

The wavefront model represents the extended emission region by an effective point source. Cross-correlation reduces relative threshold-crossing offsets, but a sparse array constrains direction more tightly than longitudinal source distance. Errors in the fitted source and axis affect response inversion, the radio-footprint centre and the radial distances used in the LDF fit.

Azimuthal averaging removes the leading charge-excess harmonic from the reference LDF, but triggered stations generally sample azimuths unevenly. A residual charge-excess contribution may therefore remain in the event amplitudes. The paired controls show an effect of reconstructed geometry, while the primary-specific means show sensitivity to reference choice. These effects motivate a two-dimensional field model and amplitude corrections calibrated using reconstructed source parameters.

Zhang et al.~\cite{refZhang2026E2E} combine spherical-wavefront and electric-field ADF fits and estimate primary energy from geomagnetic radiation energy. The main differences here are direction fitting in voltage, refracted-ray viewing angles and energy estimation from finite-band peak fields. Table~\ref{tab:method_contributions} lists the established methods and their extensions in this work.

\begin{table}[!htbp]\centering
\caption{Established methods and their applications or extensions in this reconstruction.}
\label{tab:method_contributions}
\small\renewcommand{\arraystretch}{1.15}\setlength{\tabcolsep}{4pt}
\begin{tabular}{@{}>{\raggedright\arraybackslash}p{0.17\textwidth}>{\raggedright\arraybackslash}p{0.34\textwidth}>{\raggedright\arraybackslash}p{0.44\textwidth}@{}}\toprule
Component & Established basis & Implementation and purpose in this work\\\midrule
Wavefront timing & Plane- and spherical-wavefront reconstruction~\cite{refLOPESWavefront,refDecoene2023,refZhang2026E2E}. & Correct relative L1 offsets by cross-correlation; integrate refractivity for each source--station path to account for pulse timing and propagation differences (sections~\ref{sec:timing} and~\ref{sec:wavefront}).\\\addlinespace
Particle axis & Angular-distribution fitting~\cite{refZhang2026E2E}. & Fit voltages before field inversion to obtain its required direction; use source-end ray tangents to include refraction in the viewing angles.\\\addlinespace
Electric field & Analytical three-polarisation least-squares inversion and noise weighting~\cite{refThreePolEField}; detector-response model~\cite{refNoiseRFI}. & Extend the input from open-circuit voltages to ADC traces by de-embedding the loaded RF network; evaluate signal and noise weights at the same reference plane (section~\ref{sec:rf_inverse}).\\\addlinespace
Footprint geometry & Refractive footprint displacement and early--late mapping~\cite{refSchlueter2020CoreShift,refSchlueter2023}. & Use the same fitted source and axis to determine the shifted radial origin and the station coordinates.\\\addlinespace
Reference LDF & ZHAireS electric-field simulations~\cite{refZHAireS}; reference-distribution construction in FARSim~\cite{refFARSim}. & Fit reconstructed peak amplitudes to the proton and iron references; average the two energy estimates without assigning primary mass.\\\addlinespace
Peak amplitude & Magnetic coherence-scale estimate~\cite{refChiche2024}; density-dependent energy correction~\cite{refZhang2026E2E}. & Calibrate the finite-band event-to-reference amplitude ratio to account for differing source conditions; test a Fresnel interpretation (section~\ref{sec:coherence}).\\\bottomrule
\end{tabular}\end{table}

The results are obtained within a common ZHAireS, atmospheric and detector-response model. The detector-level performance covers $60^\circ\leq\theta_{\rm sim}<85^\circ$ with the specified external and internal noise. Application to measured events requires station timing, positions, gains and complex response calibrations~\cite{refGRANDlib}.

\Needspace{14\baselineskip}
\section{Conclusion}
\label{sec:conclusion}

The reconstruction uses pulse cross-correlation and a voltage-domain ADF with refracted-ray viewing angles to determine the arrival direction from self-triggered ADC traces. The fitted source and particle axis define both the ADF geometry and the particle-core and radio-footprint-centre positions. De-embedding the loaded RF network extends the three-polarisation inversion from our previous collaborative work to ADC inputs. For energy estimation, the event-to-reference correction accounts for the dependence of finite-band peak amplitude on source conditions. A Fresnel parameterisation relates this dependence to finite-source interference.

The field-level benchmark assesses the attainable energy performance of the chosen LDF estimator with simulated fields and input geometry. Its relative energy RMSE is 11.90\% without the peak-amplitude correction and 8.31\% with either the empirical or semi-analytic parameterisation. The complete detector-level test contains 3,073 common events at $60^\circ\leq\theta_{\rm sim}<85^\circ$. Across three waveform treatments, median opening-angle errors are $0.0315^\circ$--$0.0349^\circ$, mean energy residuals are $-2.30\%$ to $-1.86\%$, standard deviations are 8.46\%--8.96\%, and RMSE values are 8.77\%--9.15\%. The empirical correction reduces the overall energy error. The remaining dependence on zenith angle and primary particle motivates tighter geometrical constraints and more detailed spatial reference models.

\appendix

\section{Paired geometry and field-inversion controls}
\label{app:audits}

The input-geometry control records cover 3,052 of the 3,073 main events: 1,471 iron and 1,581 proton showers. The other 21 events lack both controls and are excluded only from this appendix. For this diagnostic, the simulated primary selects its matching reference, rather than the equal-weight estimator used for the main performance. Table~\ref{tab:paired_geometry} compares the input field with simulation-input geometry, the ADC-recovered field with the same geometry, and the current three detector branches with reconstructed geometry.

\begin{table}[!htbp]
\centering
\caption{Paired primary-matched controls on 3,052 events. The simulated primary selects the reference only in this diagnostic. Energy residuals are percentages.}
\label{tab:paired_geometry}
\setlength{\tabcolsep}{7pt}
\renewcommand{\arraystretch}{1.15}
\begin{tabular}{llrrrr}
\toprule
Primary & Input and geometry & $N$ & Mean & SD & RMSE\\
\midrule
Iron & Input field; input geometry & 1471 & +0.30 & 6.96 & 6.96\\
Iron & Recovered field; input geometry & 1471 & +0.30 & 6.96 & 6.96\\
Iron & Noise free; reconstructed geometry & 1471 & -1.57 & 7.95 & 8.10\\
Iron & Full trace; reconstructed geometry & 1471 & -1.15 & 8.47 & 8.54\\
Iron & Trigger window; reconstructed geometry & 1471 & -1.39 & 8.35 & 8.46\\
Proton & Input field; input geometry & 1581 & -2.08 & 7.95 & 8.21\\
Proton & Recovered field; input geometry & 1581 & -2.08 & 7.95 & 8.21\\
Proton & Noise free; reconstructed geometry & 1581 & -3.31 & 8.42 & 9.05\\
Proton & Full trace; reconstructed geometry & 1581 & -2.85 & 8.92 & 9.37\\
Proton & Trigger window; reconstructed geometry & 1581 & -3.12 & 8.79 & 9.32\\
\bottomrule
\end{tabular}
\end{table}

The input-field and recovered-field mean residuals differ by less than 0.01 percentage point for either primary under the same supplied geometry. Replacing that geometry by the current noise-free reconstruction changes the iron mean from $+0.30\%$ to $-1.57\%$ and the proton mean from $-2.08\%$ to $-3.31\%$. This paired test supports a small numerical inversion bias in the matched-response model and a larger contribution from geometry and reference matching. 

\section{L1-anchored waveform gate}
\label{app:gate}

Let $\Delta k=k-k_{\rm L1}$ denote the sample offset from the recorded station trigger. The denoised waveform is $V_{\rm den}[k]=w(\Delta k)V_{\rm raw}[k]$, with
\begin{equation}
w(\Delta k)=
\begin{cases}
\frac{1}{2}\left[1-\cos\left(\pi\frac{\Delta k+36}{24}\right)\right],
& -36\leq\Delta k<-12,\\
1, & -12\leq\Delta k\leq24,\\
\frac{1}{2}\left[1+\cos\left(\pi\frac{\Delta k-24}{40}\right)\right],
& 24<\Delta k\leq64,\\
0, & \text{otherwise}.
\end{cases}
\end{equation}
Figure~\ref{fig:gate_example} illustrates the waveform gate defined here. At 2 ns sampling, the unity core spans $-24$ to $+48$ ns relative to L1 and the complete support spans $-72$ to $+128$ ns. The measured samples in the core are unchanged; only the shoulders are tapered.

\acknowledgments

This work was supported by the National Square Kilometre Array (SKA) Program of China under Grants 2025SKA0110101 and 2025SKA0110103.

\section*{Data and code availability}

Event-level tables for the 4,384 triggered events, the 3,083 events selected in all three waveform modes and the 3,073-event principal analysis sample are available from the corresponding author on reasonable request. The same request may include zenith summaries, reference-LDF and peak-amplitude-correction comparisons, source data for every reported figure and the scripts used to reproduce the tables and figures. Waveform-level simulation products, electric-field calibration traces and MAT-format reference-LDF files require a separate transfer arrangement because of their size. The LDF construction, radio-centre convention and normalisation are described in section~\ref{sec:method} and follow the FARSim framework.

\end{document}